\documentclass[12pt,a4paper]{article}
\pdfoutput=1
\usepackage{graphicx}
\usepackage{amsmath,amsfonts,amssymb,setspace}
\usepackage{slashed}
\usepackage{braket,xcolor}
\usepackage{verbatim}
\usepackage{caption}
\usepackage{subcaption}
\usepackage{multirow}
\usepackage{amsfonts}
\usepackage[utf8]{inputenc}
\usepackage{hyperref}
\usepackage{cite}

\hypersetup{colorlinks=false, linkcolor=blue, citecolor=red}

	\def\be{\begin{equation}}
	\def\ee{\end{equation}}
	\def\bea{\begin{eqnarray}}
	\def\eea{\end{eqnarray}}
	\def\ba{\begin{array}}
		\def\ea{\end{array}}
	\def\ben{\begin{enumerate}}
		\def\een{\end{enumerate}}

	\newcommand{\vs}[1]{\vspace{#1 mm}}
	\newcommand{\dsl}{\pa \kern-0.5em /}

	\def\pa {\partial}

\begin{document}
	\begin{flushright}
		%
	\end{flushright}
	\begin{center}
		{\Large{\bf Complexity from the Reduced Density Matrix:  a {\it new} Diagnostic for Chaos}}
		
		\vs{10}
		
		{Arpan Bhattacharyya${}^{a}$\footnote{abhattacharyya@iitgn.ac.in}, S. Shajidul Haque${}^{b}$\footnote{shajid.haque@uct.ac.za}, Eugene H. Kim${}^{c}$\footnote{ehkim@uwindsor.ca}} 
		
		\vskip 0.3in
		
		{\it ${}^{a}$Indian Institute of Technology, Gandhinagar, Gujarat-382355, India}
		\vskip .5mm
		{\it ${}^{b}$High Energy Physics, Cosmology \& Astrophysics Theory Group \\and The Laboratory for Quantum Gravity \& Strings,\\
Department of Mathematics and Applied Mathematics,\\
University of Cape Town, South Africa}
		\vskip .5mm
		{\it ${}^{c}$Department of Physics, University of Windsor, \\ Windsor, Ontario N9B 3P4 Canada}\vskip .5mm

		
		
		
		
		
	\end{center}

	\begin{abstract}
		We investigate circuit complexity to characterize chaos in multiparticle quantum systems. In the process, we take a stride to analyze open quantum systems by using complexity. We propose a new diagnostic of quantum chaos from complexity based on the reduced density matrix by exploring different types of quantum circuits. Through explicit calculations on a toy model of two coupled harmonic oscillators, where one or both of the oscillators are inverted, we demonstrate that the evolution of complexity is a possible diagnostic of chaos.
	\end{abstract}
	\newpage
	\tableofcontents






\section{Introduction}
Chaotic systems are abundant in nature. Although we have a reasonably thorough understanding of classical chaos, our knowledge of chaos for quantum systems is still inadequate \cite{NHJ,Jahnke:2018off}. We expect quantum chaos to play an essential role in understanding some of the most important open questions in physics, such as thermalization, transport in quantum many-body systems, black hole information loss, to name a few. Therefore, it is essential to have a comprehensive understanding of quantum chaos \cite{NHJ,Jahnke:2018off}.

Traditionally, chaos in quantum systems has been identified by comparing results from random matrix theory (RMT) \cite{Bohigas:1983er,Jahnke:2018off}. Recently, however, other diagnostics have been proposed to probe chaotic quantum systems. One such diagnostic is the out-of-time-ordered correlator (OTOC), $F_T(t) \equiv \langle W(t) V(0) W(t) V(0) \rangle,$ where $W(t)$ and $V(t)$ are operators in the Heisenberg picture, and the angle bracket $\langle \cdots \rangle$ denotes a thermal average. 
This quantity has been argued to give information about the chaos in quantum mechanical systems \cite{Kitaev2015,Larkin1969,Maldacena_2016}. It has been shown that the (classical) Lyapunov exponent and the scrambling time may be extracted from these quantities. However, recent works have revealed some tension between the OTOC and RMT diagnostics \cite{Hashimoto:2017oit}. A deeper understanding of probes of quantum chaos is required; it is worthwhile to investigate other probes of quantum chaos. Quantum information theory turns out to be the most promising in this direction.  

Ideas and tools from quantum information theory have come to permeate all of modern (quantum) physics.  They have brought new insights into traditional ideas and had far-reaching consequences, e.g. a new perspective in the structure of space-time itself.
Most information-theoretic studies begin by (purposely) partitioning the system into subsystems $A$ and $B$; one considers the reduced density matrix of subsystem-$A$, upon tracing out subsystem-$B$,
$ \hat{\rho}_A = {\rm Tr}_B[ \hat{\rho} ]$, where $\hat{\rho}$ is the density matrix of the entire system and the {\sl von Neumann entropy},
$ S = - {\rm Tr} [ \hat{\rho}_A \ln \hat{\rho}_A ] $. 
Indeed, much of quantum information theory is concerned with the uses and interpretation of the von Neumann entropy.  

An information-theoretic quantity that has recently come into the limelight is the system's {\sl complexity}. Complexity, rather `Circuit Complexity' \cite{NL1,NL2,NL3}, is an idea from the theory of quantum computation --- it is the shortest distance between some reference states $| \psi_R \rangle$ and a target state $| \psi_T \rangle$. Operationally, it quantifies the minimal number of operations 
needed to manipulate $| \psi_R \rangle$ to $| \psi_T \rangle$. 
The flurry of recent work on circuit complexity in the field of quantum field theory in recent time  \cite{Jefferson,Chapman:2017rqy,Caputa:2017yrh,Bhattacharyya:2018bbv,Hackl:2018ptj,Khan:2018rzm,Camargo:2018eof,Ali:2018aon,Bhattacharyya:2018wym,Caputa:2018kdj,Bhattacharyya:2019kvj,Caputa:2020mgb,Flory:2020eot,Erdmenger:2020sup, cosmology1,cosmology2,DiGiulio:2020hlz,Caceres:2019pgf,Susskind:2020gnl,Chen:2020nlj,Czech:2017ryf,Camargo:2019isp,Chapman:2018hou,Chapman:2019clq,Doroudiani:2019llj,Geng:2019yxo,Guo:2020dsi} has largely been spurred by black hole physics and, in particular, the conjecture that it resolves certain puzzles related to black holes \cite{susskind1,susskind2}. 
The aspect of complexity that we are particularly interested in is its potential to characterize quantum chaos.  
Previous work \cite{me3,Bhattacharyya:2019txx,Ryuchaos,IHOme,Balasubramanian:2019wgd,Yang:2019iav,Yang:2019udi} has shown that complexity could detect the scrambling time and Lyapunov exponent. However, this required a unique type of quantum circuit --- a doubly-evolved state, where the target state is obtained by first evolving the system forward in time with a Hamiltonian $H$, and then evolving it backwards in time with a slightly different Hamiltonian $H + \delta H$.


In this work, we take the first step toward using complexity to characterize chaos in a multiparticle system. We consider {\it a toy model} consisting of two coupled oscillators, where one or both of the oscillators are inverted. We will refer throughout the paper, one of the oscillator as the `\textit{system}' and the other one as the `\textit{bath}'. It is well known that, classically, the inverted oscillator has an unstable fixed point in phase space at $(x=0, p=0)$  (and, hence, not chaotic in the strict sense). Nonetheless, in the context of studying quantum chaos in various quantum field theories \cite{Blume-Kohout2003-ww,Morita2019-de,Bueno2019-zh,Yan2020-ov,Betzios:2016yaq,Betzios:2020wcv,Hegde:2018xub,Hashimoto:2020xfr}, it has served as a powerful toy model mostly because it is an exactly solvable system.
We begin by revisiting the complexity of the doubly-evolved circuit state, namely a state obtained by first evolving the system forward in time with a Hamiltonian $H$, and then evolving it backwards in time with a Hamiltonian $H + \delta H$. We highlight some new features that were not appreciated in previous works \cite{me3}. We discovered that in the inverted regime the complexity for this doubly evolved state and a single evolved state with respect to the same reference state saturate at the same value. When the parameters for the system and bath are different, the linear growth region of complexity splits into two separate regions, indicating the Lyapunov spectrum \cite{Spectrum}. 

Then we propose a new diagnostic of chaos using complexity that does not require this type of contrived target state and is, in fact, more powerful. This new diagnostic is based on the reduced density matrix by employing the  ``operator-state mapping" \cite{JAMIOLKOWSKI1972275,CHOI1975285,PhysRevA.87.022310} to build an effective target state.
Our proposal captures all the features mentioned above and more. For example, we found that the scrambling time and Lyapunov exponents are mainly dictated by the bath parameter. 
Finally, we also compare this new form of complexity with the complexity of purification \cite{Agn2019,Caceres:2019pgf,Camargo:2018eof,Camargo:2020yfv,Bhattacharya:2019zkb,Chen:2018mcc} for detecting chaos. We discover that the complexity of purification is not as sensitive as our proposed probe. 

It should be noted, the model we consider can be thought of as the simplest example of an {\sl open quantum system}, where some subsystem is treated as the `system', and the rest is treated as a `bath/reservoir' \cite{Blume-Kohout2003-ww}. Open quantum systems are extremely important for various reasons 
\cite{PhysRevLett.46.211, CALDEIRA1983374,open}. Our proposal takes a step toward using complexity to characterize open quantum systems. Previously in \cite{cosmology1,cosmology2}, a system connected with a classical source was considered; however, the diagnostic used was the full system complexity.  

The rest of the paper is organized as follows. In Section 2, we present our model and a brief review of circuit complexity. In Section 3, we compute the evolution of complexity for the entire system for different circuits and demonstrate how quantum chaos can be detected and quantified. In Section 4, we illustrate our new proposal as the diagnostic of chaos and compare it with the complexity of purification. Finally, we summarize and present concluding remarks in Section 5.


\section{Our Model and Complexity}
Our model consists of two oscillators, where we treat one of the oscillators as the {\bf system} and the other as the {\bf bath}. The Hamiltonian of our model is the following
\begin{equation}
 H = \frac{1}{2} p_1^2 + \frac{\omega_0^2 \epsilon_1}{2} x_1^2
    + \frac{1}{2} p_2^2 + \frac{\omega_0^2 \epsilon_2}{2} x_2^2
    + \lambda  \omega_0^2  x_1 x_2  \ .
\label{Hsystem}
\end{equation}
Here $x_i$ and $p_i$ are the position and momentum operators at site-$i$, with $[x_i, p_j] = i \delta_{i,j}$, $\omega_0$ is a parameter with units of energy and $\{\epsilon_1,\epsilon_2,\lambda\} \in {\bf R}$ are dimensionless parameters.  
In what follows, we take $\lambda \geq 0$; we set $\omega_0=1$, i.e. $\omega_0$ sets the energy scale.
We are working in units where $\hbar = 1$.
The Hamiltonian (Eq.~\ref{Hsystem}) is readily diagonalized by introducing a matrix notation --- 
\begin{subequations}
\begin{equation}
 {\bf x} =  \left( \begin{array}{c}
    x_1 \\ x_2 
    \end{array} \right) 
 \ \ , \ \ 
 {\bf p} =  \left( \begin{array}{c}
    p_1 \\ p_2 
    \end{array} \right) 
 \ \ ,  \ \
 \hat{K} = \left( \begin{array}{cc}
    \epsilon_0 & \lambda \\ 
    \lambda & -\epsilon_0
    \end{array} \right)   \ ,
\end{equation}
where $\epsilon_0 = (\epsilon_1 - \epsilon_2)/2$. Using this matrix notation, the Hamiltonian takes the following form
\begin{equation}
 H = \frac{1}{2} {\bf p}^T {\bf p} + \frac{\omega_0^2}{2} \left (
        \frac{1}{2}(\epsilon_1 + \epsilon_2) {\bf x}^T {\bf x}
      + {\bf x}^T \hat{K} {\bf x} \right ) \ .
\label{Hmatrix}
\end{equation}
\end{subequations}
By performing an orthogonal transformation, we can diagonalize this Hamiltonian (\ref{Hmatrix}). Under this transformation we define
new position and momentum variables $Q_i$ and $P_i$ ($[Q_i,P_j] = i \delta_{i,j}$) as
\begin{equation}
 {\bf x} = S {\bf Q} \ , \ \ {\bf p} = S {\bf P},
 \label{transformation}
 \end{equation}
where
\begin{equation} 
 {\bf Q} =  \left( \begin{array}{c}
    Q_s \\ Q_a 
    \end{array} \right) 
 \ \ , \ \ 
 {\bf P} =  \left( \begin{array}{c}
    P_s \\ P_a 
    \end{array} \right) 
 \ \   \ \ \text{and} \ \
 S =  \left( \begin{array}{cc}
    u & -v \\ 
    v & u
    \end{array} \right)
 \end{equation}
with
\begin{equation}
  u = \sqrt{ \frac{1}{2} \left( 1 + \frac{\epsilon_0}{E} \right) }
 \ , \ 
  v = \sqrt{ \frac{1}{2} \left( 1 - \frac{\epsilon_0}{E} \right) }  \ ,
\ \ {\rm and} \ \  u^2 + v^2 = 1, \    E = \sqrt{\epsilon_0^2 + \lambda^2}. \label{eqqqq}
\end{equation}
Then Eq.~\ref{Hmatrix} takes the form
\begin{equation}
 H = \frac{1}{2} {\bf P}^T {\bf P} + \frac{\omega_0^2}{2} \left (
        \frac{\epsilon_1 + \epsilon_2}{2} \ {\bf Q}^T {\bf Q}
      + {\bf Q}^T \Lambda {\bf Q} \right )
\end{equation}
 where
\begin{equation}
 \Lambda =  \left( \begin{array}{cc}
    E & 0 \\ 
    0 & -E
    \end{array} \right)  \ .
\nonumber
\end{equation}
We can write this explicitly as follows:
\begin{equation}
 H = \frac{1}{2} P_s^2 + \frac{\Omega_s^2}{2} Q_s^2
    + \frac{1}{2} P_a^2 + \frac{\Omega_a^2}{2} Q_a^2 ,
    \end{equation}   
 where
 \begin{equation}
 \Omega_s^2 = \omega_0^2 \left (  \frac{\epsilon_1 + \epsilon_2}{2} + E \right)
 \ \ , \ \ 
 \Omega_a^2 = \omega_0^2 \left (\frac{ \epsilon_1 + \epsilon_2}{2} - E \right) \ .
\label{Hdiagonal}
\end{equation} 
The states we will consider originate from a quench in the above model --- 
\begin{eqnarray} 
 H & = & H_<  : \{\epsilon_1 = 1 \ , \ \epsilon_2 = 1 \ , \ \lambda = 0\} \ \ \ \ \ \ \ \ \ \ \  {\rm for} \ t < 0
  \\ 
 H & = & H_>  : \{ \epsilon_1 \neq 1 \ , \ \epsilon_2 \neq 1 \ , \ \lambda > 0  \} \ \ \ \ \ \ \ \ \ \ \  {\rm for} \ t > 0  \ ,
\label{newb}
\end{eqnarray}

Now we will give a brief review of circuit complexity. We will directly use the wavefunction and compute the circuit complexity using Nielsen's method \cite{NL1,NL2,NL3}. The details can be found in \cite{Jefferson}. The problem/goal of complexity is the following: given a set of elementary gates and a reference state, what is the most efficient quantum circuit that starts at that reference state (at $s=0$) and terminates at a target state ($s=1$)
\begin{equation}
    |\Psi_{s=1}\rangle = U (s=1) |\Psi_{s=0}\rangle,
\end{equation}

where $U$ is the unitary operator that takes the reference state to the target state.  We will represent the target sate $|\Psi_{s=1}\rangle$ as $|\Psi_T\rangle$ and the reference state $|\Psi_{s=0}\rangle$ as $|\Psi_R\rangle$ in the rest of the paper. We construct it from a continuous sequence of parametrized path ordered exponential of a  \textit{control} Hamiltonian operator 
\begin{equation}
U(s)= {\overleftarrow{\mathcal{P}}} \exp[- i \int_0^{s} \hspace{-0.1in} ds' H(s') ] \ .
\end{equation}
Here $s$ parametrizes a path in the space of the unitaries and given a set of elementary gates $M_I$, the  \textit{control} Hamiltonian can be written as
\begin{equation}
    H(s)= Y^{I}(s) M_{I}\, .
\end{equation} 
The coefficients $Y^I$ are the control functions that dictates which gate will act at a given value of the parameter. The control function is basically a tangent vector in the space of unitaries and satisfy the Schrodinger equation \footnote{From Eq.~\ref{invert} we have $Y^I(s) M_I= \frac{d U(s)}{ds}. U^{-1
}(s).$ $M_I's$ are taken to be the generators of some groups and can be normalized in a way that they satisfy $M_I M_J^T=\delta_{IJ}.$ Then finally we have, $ Y^{I}(s)= \textrm{Tr} \Big(\frac{d U(s)}{ds}. U^{-1
}(s). M_I^{T}\Big).$ Here $T$ denotes the transpose of matrix. For further details interested readers are referred to \cite{Jefferson}.}
\begin{equation} \label{invert}
\frac{d U(s)}{ds} = -i\, Y^I(s) M_I U(s)\,.
\end{equation}
Then we define a cost functional $\mathcal{F} (U, \dot U)$ as follows \footnote{The dot defines the derivative w.r.t s.}:
\begin{equation}
{\mathcal C}(U)= \int_0^1 \mathcal{F} (U, \dot U) ds\, .
\end{equation}
Minimizing this cost functional gives us the optimal circuit. There are different choices for the cost functional \cite{Jefferson}. In this paper we will consider
\begin{equation}
\mathcal{F}_2 (U, Y)  = \sqrt{\sum_I (Y^I)^2}\, .
\label{quadCost}
\end{equation}
In this work we will consider the ground state of $H_<$, $ | \psi_0 \rangle$ as our reference state $| \psi_R \rangle$ and we will consider different target states $| \psi_T \rangle$ that are evolved from this ground state. The target wave function 
$\psi_T( x_{1}, x_{2} ) = \langle x_1, x_2 | \psi_T \rangle$ takes the following form
\begin{equation}
 \psi_T( x_{1}, x_{2} ) = {\cal N}(t)
     \exp \left ( - \frac{1}{2} \left[ \Omega_{1}(t) x_{1}^2 + \Omega_{2}(t) x_{2}^2 
                    - 2\,\kappa(t) x_1 x_2 \right]  \right )  \ ,
\label{wavefunction}
\end{equation} 
where $\Omega_{1}(t)$, $\Omega_{2}(t) \in {\bf C}$ 
(with ${\rm Re}[\Omega_{1}(t)]$, ${\rm Re}[\Omega_{2}(t)] > 0$) 
and $\kappa(t)$, ${\cal N}(t) \in {\bf R}$. 
Following \cite{me1}, we will take the elementary gates $M_I$ as the generators of the $GL (2, C)$ group. For details interested readers are referred to \cite{me1}. In all the cases, the complexity takes the form  \cite{me1} (due to the structure of the wave functions)
\be
\mathcal{C} =\frac{1}{2} \sqrt{ \sum_{\alpha = 1,2} \left (
     \ln \left (| \hat \Omega_{\alpha}| \right)^2 
  + \tan^{-1} \left (-\frac{ \text{Im} ( \hat \Omega_{\alpha})}{\text{Re} ( \hat \Omega_{\alpha})} \right)^2 
  \right ) }  \ ,
  \label{complexity}
\ee
 where $\alpha=1, 2$ and the normal mode frequencies are given by,
\begin{align}
\begin{split}
&\hat \Omega_1=\frac{1}{2}\Big(\Omega_1(t)+\Omega_2(t)+\sqrt{(\Omega_1(t)-\Omega_2(t))^2+ 4\kappa(t)^2}\Big), \\&\hat \Omega_2=\frac{1}{2}\Big(\Omega_1(t)+\Omega_2(t)-\sqrt{(\Omega_1(t)-\Omega_2(t))^2+ 4\kappa(t)^2}\Big).
\end{split}
\end{align}

In this work, we start with first the full system complexity and study the unstable behaviour by using the techniques mentioned in \cite{me3}. 
Then we will propose a new diagnostic for studying chaos from the reduced density matrix.  


\section{System Complexity}
In this section, we discuss the complexity of the (entire) system. 
We consider two types of target states and study the evolution of complexity for the same reference state. 

\noindent
{\bf Target state I:} First we consider a target state obtained by evolving the system forward in time by the Hamiltonian $H_>$ 
\begin{equation} | \psi(t) \rangle = \exp(-i H_> t) | \psi_0 \rangle. 
\label{psiF}
\end{equation}
 $|\psi_0\rangle$ is the ground state of $H_<$ defined in Eq. \ref{newb}.
Working in the position representation, the evolved wavefunction can be written as 
\begin{subequations}
\begin{equation}
 | \psi(t) \rangle = \exp(-i H_> t) | \psi_0 \rangle 
 \longrightarrow
 \psi({\bf x},t) = \int d{\bf x}'~  K({\bf x},t | {\bf x}', t=0)~ \psi_0({\bf x}') \ ,
\end{equation}
where the initial wave function $\psi_0({\bf x})$ has the form
\begin{equation}
 \psi_0({\bf x}) = \left ( \frac{ {\rm det}( \hat{\Omega}_0 ) }{ \pi^2 } \right)^{1/4}
  \exp \left ( - \frac{1}{2} {\bf x}^T \hat{\Omega}_0 {\bf x} \right ) \ .
\end{equation}
Here $\hat \Omega_0=\omega_0$ (defined in Eq. \ref{Hsystem}) and  $K({\bf x}, {\bf x}' | t )$ is the propagator
\begin{equation}
 K({\bf x}, {\bf x}' | t ) = \langle {\bf x} | \exp(-i H_> t) | {\bf x}' \rangle
 \ .
\end{equation}
\end{subequations}
For the Hamiltonian given in Eq.~\ref{Hsystem}, the propagator factorizes in normal mode coordinates and the propagator for each normal mode is 
\begin{equation}
 K(Q_{\alpha}, Q'_{\alpha} | t ) 
  = \left( \frac{g_{\alpha}}{i 2\pi } \right)^{1/2}
     \exp \left ( \frac{i}{2} \left[ 
       f_{\alpha} (Q_{\alpha}^2 + Q_{\alpha}'^2)  
   - 2 g_{\alpha} Q_{\alpha} Q'_{\alpha}  \right] \right )
   \end{equation}
 where
$ f_{\alpha} = \Omega_{\alpha} \cot (\Omega_{\alpha} t),$ and $ g_{\alpha} = \Omega_{\alpha}/\sin (\Omega_{\alpha} t)$ with $\Omega_{\alpha}'s$ are defined via Eq. \ref{Hdiagonal} and $\alpha=a, s$.
%
To carry out the calculations, we return to the original coordinates via Eq.~\ref{transformation}. The propagator can be written as
\begin{subequations}
\begin{equation}
  K({\bf x}, {\bf x}' | t ) = \prod_{\alpha} \left( \frac{g_{\alpha}}{i 2\pi } \right)^{1/2} 
    \exp \left ( \frac{i}{2} \left[ 
     ( {\bf x}^T \hat{F} {\bf x} + {\bf x}'^T \hat{F} {\bf x}' ) 
  - {\bf x} \hat{G} {\bf x}' - {\bf x}' \hat{G} {\bf x}  \right] \right )  \ ,
\label{propagatormatrix}
\end{equation}
where
\begin{equation}
 \hat{F} = S  
  \left( \begin{array}{cc}
    f_S & 0 \\ 
    0 & f_A
    \end{array} \right) S^T
 \ \ , \ \ 
 \hat{G} = S  
  \left( \begin{array}{cc}
    g_S & 0 \\ 
    0 & g_A
    \end{array} \right) S^T
 \ .
\end{equation}
\end{subequations}
Carrying out the Gaussian integrals, one obtains
\begin{equation}
 \psi({\bf x}, t) = \left ( \frac{ {\rm det}( {\rm Re}[\hat{\Omega}(t)] ) }{\pi^2} \right )^{1/4}
  \exp \left( - \frac{1}{2} {\bf x}^T \hat{\Omega}(t) {\bf x} \right)  \ ,
\label{wavefunctionmatrix}
\end{equation}
where
\begin{equation} 
\hat{\Omega}(t) = \hat{G} ( \hat{\Omega}_0 - i \hat{F} )^{-1} \hat{G} - i \hat{F} 
 \ .
\end{equation}
{\bf Target state II:} To get the other target state we evolve the system forward in time with a Hamiltonian $H_>^F$, and then backward in time with a slightly different Hamiltonian $H_>^B$: 
\begin{equation} 
| \psi(t) \rangle = \exp( +i H_>^B t) \exp( -i H_>^F t) | \psi_0 \rangle.
\label{psiFB}
\end{equation}
 Both $H_>^F$ and $H_>^B$ have the same form but have slightly different values of $\epsilon_1,\epsilon_2,\lambda$ defined in Eq. \ref{Hsystem}. For this case we can write in the position representation:
\bea
 | \psi(t_2) \rangle &=& \exp(-i H_2 t_2) \exp(-i H_1 t_1) | \psi_0 \rangle, \cr
 \psi({\bf x},t_2) &=& \int d{\bf x}' d{\bf x}''~  
   K_2({\bf x}, {\bf x} | t_2)~K_1({\bf x}', {\bf x}'' | t_1)~ \psi_0({\bf x}'') \ ,
\eea
where we are ultimately interested in $t_1 \rightarrow t$ and $t_2 \rightarrow -t$. Using the parameterization for the propagator in Eq.~\ref{propagatormatrix} and carrying out the (Gaussian) integrals, one obtains Eq.~\ref{wavefunctionmatrix} with
\begin{equation}
 \hat{\Omega}(t) = \hat{G}_2 \hat{M}_2^{-1} \hat{G}_2 - i \hat{F}_2 
 \ \ {\rm with} \ \
  \hat{M}_2 = \hat{G}_1 ( \hat{\Omega_0} - i \hat{F}_1 )^{-1} \hat{G}_1 - i ( \hat{F}_1 + \hat{F}_2 ) 
\ .\label{eqq}
\end{equation}
Then the wavefunction can be written as the desired form mentioned in Eq. \ref{complexity}:
\begin{equation}
 \psi( x_{1}, x_{2} ) = \left ( \frac{ {\rm det}( {\rm Re}[\hat{\Omega}(t)] ) }{\pi^2} \right)^{1/4}
     \exp \left ( - \frac{1}{2} \left[ \Omega_{1}(t) x_{1}^2 + \Omega_{2}(t) x_{2}^2 
                    - 2\,\kappa(t) x_1 x_2 \right]  \right )  \ ,
\label{wavefunction2}
\end{equation} 
where 
\begin{equation}
 \Omega_1(t) = \hat{\Omega}(t)_{11}  \ \ , \ \
 \Omega_2(t) = \hat{\Omega}(t)_{22}  \ \ , \ \ 
 \kappa(t) = - \hat{\Omega}(t)_{12},
\end{equation}
$\hat{\Omega}(t)_{i j} $ denotes various components of the $\hat \Omega(t)$ defined in Eq. \ref{eqq}. Finally using Eq. \ref{eqqqq} and Eq. \ref{Hdiagonal} these can be written in terms of the parameters of the Hamiltonian. \par 

\begin{figure}[ht]
\begin{center}
\scalebox{.80}{\includegraphics{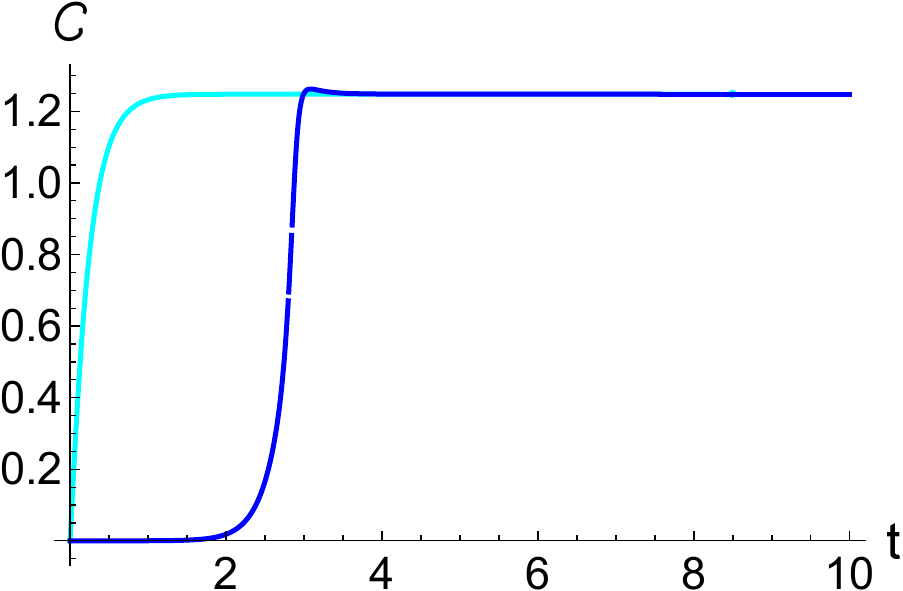} }
\scalebox{.80}{\includegraphics{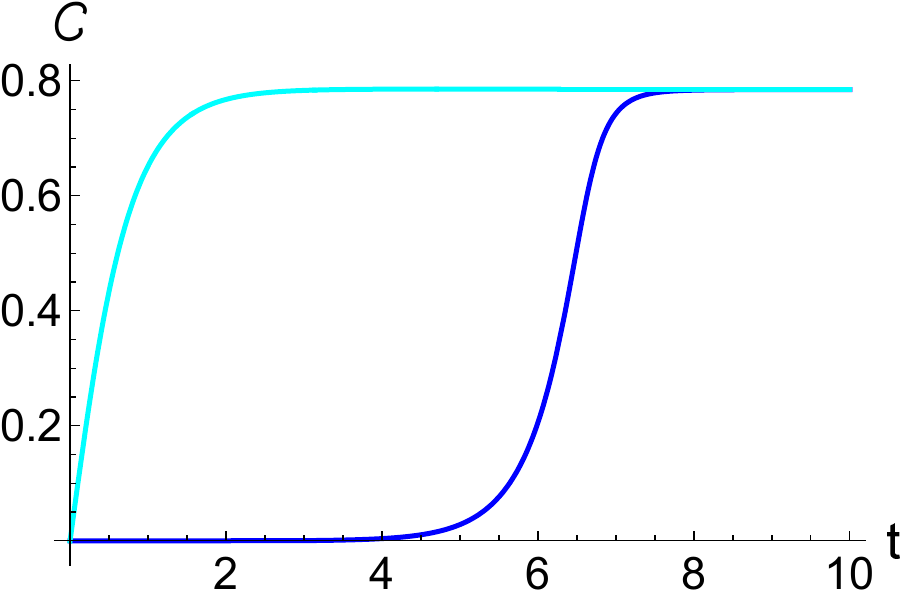} }
\vspace{-0.1in}
\caption{{\bf Left:} Complexity for single evolved (in cyan) and doubly evolved (in blue) with $\epsilon_1=-5, \epsilon_2=-5, \lambda=0.1, \delta \lambda =0.001$\, and  {\bf Right:} Complexity for single evolved (in cyan) and doubly evolved (in blue) with $\epsilon_1=-1, \epsilon_2=1, \lambda=0.1, \delta \lambda =0.001.$}
\label{LEvsFull}
\end{center}
\end{figure}

For a chaotic quantum system, even if the two Hamiltonians are arbitrarily close\footnote{For our case, this means the values of $\epsilon_1,\epsilon_2,\lambda$'s are close to each other}, i.e. $H_>^B = H_>^F + \delta H_>^F$,  the resulting state $|\psi_T\rangle$ will be quite different from $|\psi_0\rangle.$ This type of evolved state is used to compute the Loschmidt echo \cite{Gorin_2006,2012arXiv1206.6348G,Nizami:2020agu}, which is a measure of quantum chaos. It was pointed in \cite{me3} that the complexity for this type of doubly-evolved state defined in Eq. \ref{wavefunction2} is essential to capture the chaotic behaviour of the system. The authors showed that complexity could capture similar information contained in the out-of-time order correlators and, hence, can be used as an alternative diagnostic to study chaotic features. In \cite{me3}, it was shown that the \textit{time scale when complexity starts to grow linearly is equivalent to the scrambling time, and the slope of the linear portion captures the Lyapunov exponent.}

In what follows, we employ the approach of \cite{me3} and consider the complexity associated with Eq.~\ref{psiFB}, comparing it with the complexity associated with Eq.~\ref{psiF}; in doing so, we demonstrate a new feature of many-body unstable systems. We also demonstrate that the complexity associated with Eq.~\ref{psiFB} is capable of sensing the system's Lyapunov spectrum. As highlighted above, since we are interested in chaotic behaviour, we focus on the case where
either one of the oscillators or both are inverted ($\epsilon_1 <0,$ and/or $ \epsilon_2<0$). Note that there are different ways to construct the perturbed Hamiltonian. For our simple model, we have three parameters: $\{\epsilon_1, \epsilon_2$ $\lambda\}$; we can construct the perturbed Hamiltonian $H+\delta H$ by considering perturbations of any combination of these parameters. 
Fig.~\ref{LEvsFull} displays the evolution of complexity for the above mentioned target states, where $H+\delta H = H(\lambda+\delta \lambda)$. Notice that for both cases, the complexity is bounded. This behaviour was not seen in \cite{me3}; we believe this is due to the more sensitive nature of wave function method over the correlation matrix method  as discovered in \cite{me1}\footnote{ This speculation is motivated by the fact that the full system complexity is very similar to a single oscillator model used in \cite{me3}, and the main difference between these two models is the use of two different methods of complexity. Note that in \cite{me3} the complexity was computed using the covariance matrix. Since the states under consideration are Gaussian states they can be equivalently described by their covariance matrix. One can compute the complexity from the covariance matrix \cite{me1}. Also, \cite{me1} tells us that the covariance matrix method is less sensitive compared to the wavefunction method (of computing circuit complexity) as it uses different types of quantum gates for evolved states.}. \\\\ The key observations that follows from Fig.~\ref{LEvsFull} are given as follows:
\begin{itemize}
    \item The complexity for the singly-evolved state (Eq.~\ref{psiF}) grows rapidly and reaches saturation. On the other hand the complexity for the doubly-evolved state (Eq.~\ref{psiFB}) starts with a scrambling regime, grows linearly, and finally reaches its (maximum) saturation value. 
    \item  The complexity is bounded by the same values for both the singly-evolved and doubly-evolved states. 
\end{itemize}
Note that, for a regular system, the magnitude of the complexity increases when we evolve the system further. Therefore, this bounded nature of the complexity as shown in Fig.~(\ref{LEvsFull}), appears only when one or both oscillators of the system is (are) %
inverted with fixed parameters.\par

Furthermore, a state obtained by evolving a reference state multiple times will have the same complexity after the scrambling time in the inverted regime. Hence, it appears that, after a certain time scale, no target state is any more or less difficult to construct (from the same reference state) than any other. This might have potential practical applications in information processing.

\begin{figure}[ht]
\begin{center}
\scalebox{0.87}{\includegraphics{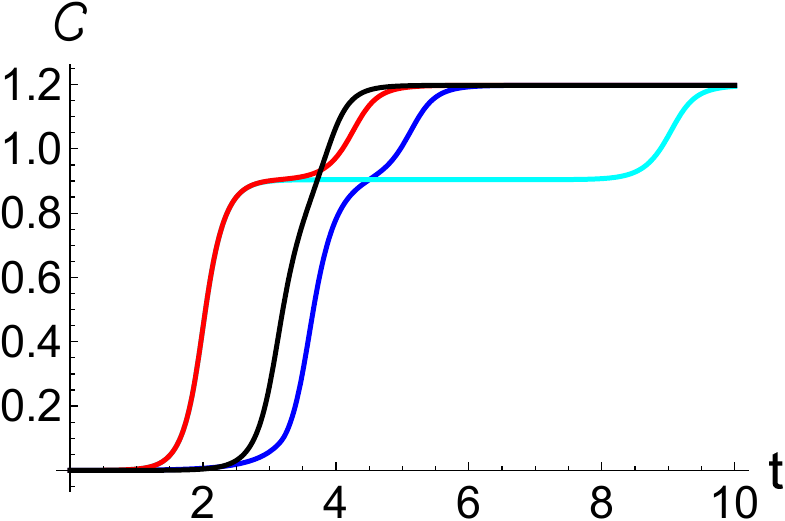} }
\scalebox{0.80}{\includegraphics{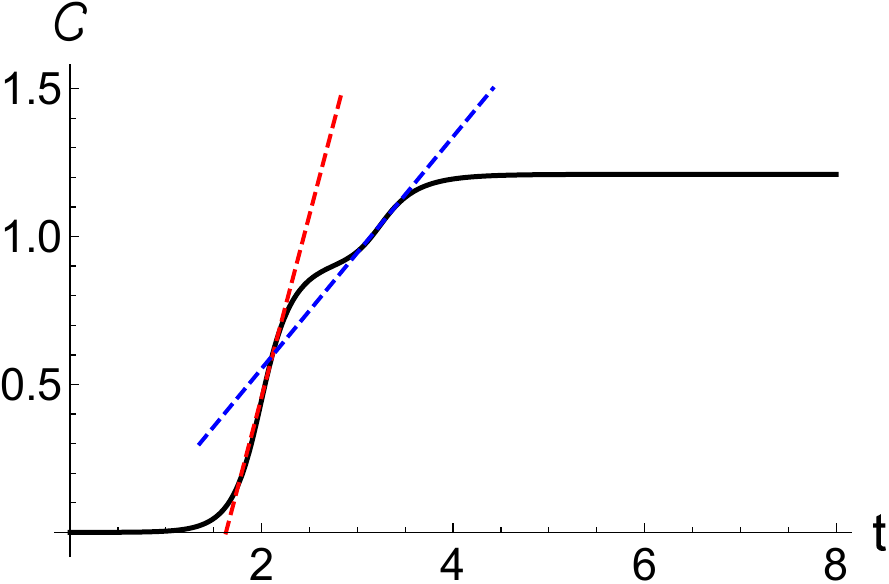} }
\vspace{-0.1in}
\caption{{\bf Left:} Complexity for forward-backward evolved target state for perturbation in different directions with $\epsilon_1= -6, \epsilon_2=-1, \lambda=0.1$ Cyan: $\delta\epsilon_1=0.001, \delta\epsilon_2=0, \delta\lambda=0$, Blue: $\delta\epsilon_1=0, \delta\epsilon_2=0.001, \delta\lambda=0$, Black: $\delta\epsilon_1=0, \delta\epsilon_2=0, \delta\lambda=0.001$, Red: $\delta\epsilon_1=0.001, \delta\epsilon_2=0.001, \delta\lambda=0$. \ \ {\bf Right:} Complexity display different Lyapunov when $\epsilon$ parameters for the system and bath oscillators are different. Here we took $\epsilon_1=-6, \epsilon_2=-2, \lambda=0.1, \delta \epsilon=0.001, \delta \lambda=0.001$ The two slopes are displayed by red and blue dashed lines.  }
\label{all}
\end{center}
\end{figure}

Note that we can perform the backward evolution with the Hamiltonian,
$H + \delta H = H(\epsilon_1+\delta \epsilon_1, \epsilon_2+\delta \epsilon_2, \lambda)$ i.e. we can perturb the Hamiltonian by slightly changing the parameters $\epsilon_1$ and $\epsilon_2$.  
Fig.~\ref{all} displays how such a perturbation in different direction displays different complexity growth, though the qualitative feature of the growth is similar.

\textit{The evolution of complexity (coming from Eq.~\ref{psiFB}) for this type of perturbation demonstrates another interesting feature of our unstable system --- when the parameters $\epsilon$'s are quite different, the linear growth of complexity happens in two distinct segments with two different slopes. The right panel of Fig.\ref{all} shows that complexity is almost flat early time (scrambling), then it grows linearly; after that it flattens again (2nd scrambling) for a period, followed by a second linear growth and finally reaching saturation.}

\begin{figure}[ht]
\begin{center}
\scalebox{0.8}{\includegraphics{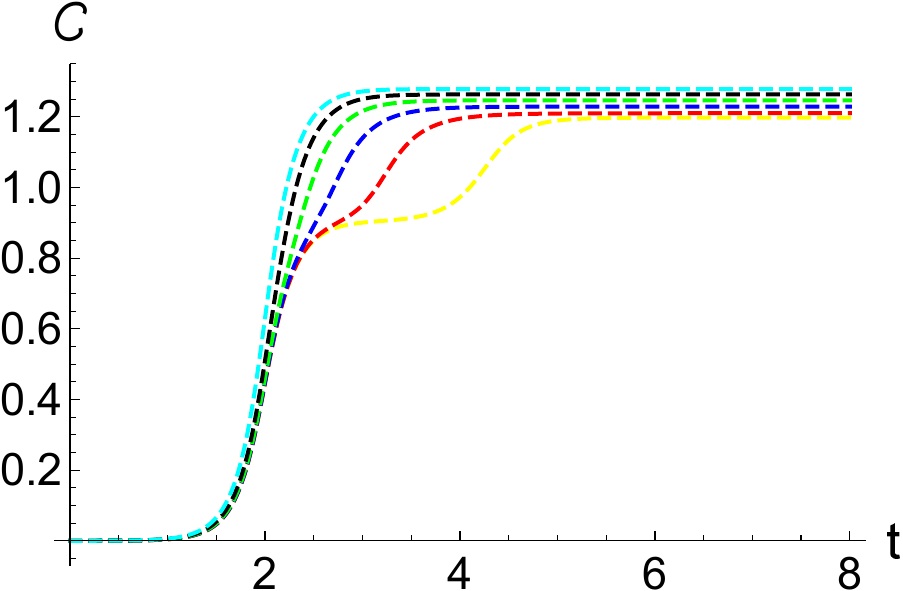} }
\scalebox{0.8}{\includegraphics{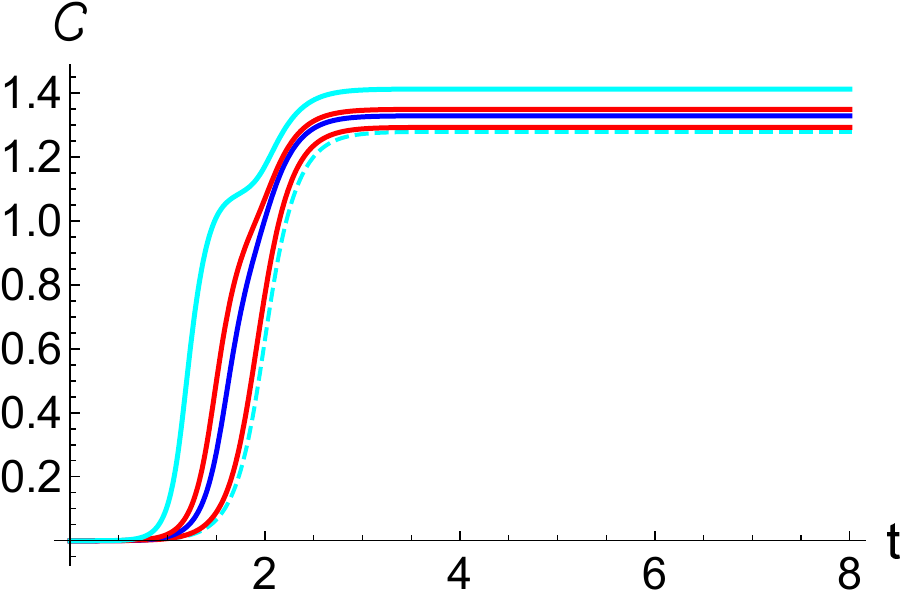} }
\vspace{-0.1in}
\caption{Complexity display two separate linear growths when the parameters ($\epsilon_1, \epsilon_2$) for the system and bath oscillators are different. In the left panel the bath parameter $\epsilon_2$ is fixed at -6 and the system parameter $\epsilon_1$ is scanning from -1 to -6. Whereas in the right panel, the system parameter $\epsilon_1$ is chosen as -6, -7, -10, -12 and -20 from dashed cyan to solid cyan curve respectively. }
\label{full2}
\end{center}
\end{figure}

To make the connection more concrete, in Fig.~\ref{full2} we fix either $\epsilon_1$ or $\epsilon_2$ and sweep through the other. We can draw some interesting conclusions, which we list below.
\begin{itemize}
\item The first linear growth and the first scrambling time is dictated by the fixed-parameter $\epsilon_2$. Fig.~\ref{full2} displays the complexity for same $\epsilon_2<0$, for different choices of $\epsilon_1<0$. 
\item
As we increase $|\epsilon_2|$, the second scrambling time decreases and the slope of the second linear portion starts to align with the slope of the first linear portion as $\epsilon_1 \rightarrow \epsilon_2$. 
\item 
When the system parameter ($|\epsilon_1|$) is bigger than the bath parameter ($|\epsilon_2|$), the slope of the upper linear portion starts to bent in the opposite direction (see Fig. \ref{full2}). 
\item 
We see the same behaviour as mentioned in the previous point, when we fix the system parameter and scan through the bath parameter. This implies that complexity for this target state, cannot distinguish between the two. 
\end{itemize}
In the next section, we propose a new diagnostic that will be able to distinguish between the two. We conclude this section by commenting on the saturation value of the complexity when $H + \delta H = H(\epsilon_1+\delta \epsilon_1, \epsilon_2+\delta \epsilon_2, \lambda)$. If we increase the magnitude of either $\epsilon_1$ and $\epsilon_2$ or both the saturation value of complexity increases. 


\section{Complexity from the Reduced Density Matrix}
In this section, we will propose a new diagnostic of chaos from complexity. First of all, this particular approach does not require the construction of the target state by performing two evolutions (forward followed by a backward with slightly different Hamiltonian) as discussed in the previous section. In that sense, this is a more natural diagnostic for studying chaotic behaviour. Secondly, it will capture all the features displayed in the previous section as well as some new ones. As noted above, our model can be thought of as the simplest open quantum system, where one of the oscillators is treated as the {\sl system} and the other is treated as the {\sl bath}. The reduced density matrix plus the operator-state mapping \cite{JAMIOLKOWSKI1972275,CHOI1975285,PhysRevA.87.022310} form the basic ingredients for constructing the diagnostic \footnote{One may be tempted to compute operator complexity for the reduced density matrix. Note that the reduced density matrix is not a unitary operator. So there is no unitary circuit which can connect it with the identity operator. It will be interesting to modify the existing methods of computing operator complexity to include non-unitary operators.} In the second part of this section, we will compare the results obtained from this approach with those obtained using the {\sl complexity of purification}, showing that the complexity of purification captures less information.

\subsection{Complexity from Operator State Mapping}

We will be interested in analyzing the reduced density matrix. 
To that end, we partition the system into two subsystems, taking oscillator-1 to be 
our ``system" and oscillator-2 to be the ``bath";
we form the reduced density matrix of oscillator-1 
\begin{equation}
 \hat{\rho}_1 = {\rm Tr}_2[ \hat{\rho} ] \ , 
\end{equation}
where $\hat{\rho}$ is the density matrix of the entire system.

Calculations are readily executed in the position representation ---
\begin{subequations}
\begin{equation}
 \hat{\rho}_1 = \int dx_1 dx_1'~ \rho_1(x_1,x_1') ~ | x_1 \rangle \langle x'_1 |  \ ,
\end{equation}
where
\begin{equation}
 \rho_1(x_1,x_1') = \int dq_2~ \rho( x_1,q_2 \mid x_1',q_2) 
\label{rhoreduced}
\end{equation}
with
\begin{equation} 
\rho(x_1,x_2 \mid x_1',x_2') = \psi(x_1,x_2) \psi^*(x_1',x_2')
\label{2particlerho}
\end{equation}
\end{subequations}
being the position-space density matrix of the full system. 
The wave function is given by Eq.~\ref{wavefunction2}, namely
\begin{equation}
 \psi( x_{1}, x_{2} ) = 
 \left( \frac{ {\rm Re}[\Omega_{s}(t)] {\rm Re}[\Omega_{a}(t)] }{\pi^2} \right)^{1/4} \exp \left ( - \frac{1}{2} \left[ \Omega_{1}(t) x_{1}^2 + \Omega_{2}(t) x_{2}^2 - 2\,\kappa(t) x_1 x_2 \right]  \right )  \ ;
\nonumber
\end{equation} 
forming the system's position-space density matrix (Eq.~\ref{2particlerho}) and tracing out oscillator-2 (as per Eq.~\ref{rhoreduced}), one obtains the reduced density matrix for oscillator-1
\begin{subequations}
\begin{equation}
 \rho_1(x_1,x_1') = \sqrt{ \frac{\gamma_1 - \eta}{\pi} }
 \exp \left ( - \frac{1}{2} \left( \gamma x_1^2 + \gamma^* x_1'^2 \right) + \eta \, x_1 x_1' \right )  \ ,
\label{rho1}
\end{equation}
where 
\begin{equation}
 \gamma = \Omega_1(t) - \kappa^2(t)/2{\rm Re}[\Omega_2(t)]\ , \ \ 
 \eta = |\kappa(t)|^2/2{\rm Re}[\Omega_2(t)]  \ ,
\end{equation}
\end{subequations}
and $\gamma_1 = {\rm Re}[\gamma]$.
In what follows, we will be particularly interested in powers of the reduced density matrix; we focus on 
$\hat{\rho}_1^{1/2}$ due to the structure of Eq.~\ref{2particlerho}, \footnote{ It will be interesting to consider other powers of $\hat \rho_1$ and check if any additional feature emerges. We are currently investigating it and hope to report about in some other publication.}
\begin{equation}
\rho^{1/2}_1(x,x') = \left( \frac{\gamma_1^2 - \eta^2}{\pi^2} \right)^{1/4}\exp \left (- \frac{1}{2} (\gamma + \eta) x^2 - \frac{1}{2} (\gamma^* + \eta) x'^2 
 + \sqrt{2 \eta (\gamma_1 + \eta) } x x' \right )\,.
\end{equation}

Now we are ready to use the operator-state mapping \cite{JAMIOLKOWSKI1972275,CHOI1975285,PhysRevA.87.022310}. The idea is for any operator, one can associate a state by working in a doubled Hilbert space.  More explicitly, for an operator $\hat{O}$, whose matrix representation with respect to the orthonormal basis $\{ | m \rangle \}$ is given by
\begin{equation}
 \hat{O} = \sum_{m,n} \hat{O}_{m,n} | m \rangle \langle n |
 \ \ {\rm where} \ \  \hat{O}_{m,n} = \langle m | \hat{O} | n \rangle \ ,
\nonumber
\end{equation}
one can associate a state with this operator via \cite{JAMIOLKOWSKI1972275,CHOI1975285} 
\begin{equation}
 \hat{O} = \sum_{m,n} \hat{O}_{m,n} | m \rangle \langle n |
 \ \  \longleftrightarrow  \ \ 
 | \hat{O} \rangle = \frac{1}{ \sqrt{{\rm Tr}[ \hat{O}^{\dagger} \hat{O} ] } }
   \sum_{m,n} \hat{O}_{n,m} | m \rangle_{\rm in} \otimes | n \rangle_{\rm out}  \ .
\end{equation}
Motivated by the thermofield-double state \cite{Hosur:2015ylk}, we work with $\hat{\rho}^{1/2}_A$; working in the position representation, one has an effective wave function (in this doubled Hilbert space)
\begin{equation}
 \psi(x,x') =    \frac{1}{ \sqrt{ {\rm Tr}[ ( \hat{\rho}^{1/2}_A )^{\dagger} 
      \hat{\rho}^{1/2}_A ] } }  \rho^{1/2}_A(x',x)  \ . 
\label{effectivepsi}
\end{equation}
We will use this as the target state and compute complexity. We can write the wavefunction explicitly as follows:
\begin{eqnarray} 
 \psi(x_1',x_1) & = & {\cal N}
   \exp \left (- \frac{1}{2} (\gamma + \eta) x_1^2 - \frac{1}{2} (\gamma^* + \eta) x_1'^2 
          + \sqrt{2 \eta (\gamma_1 + \eta) } x_1 x'_1 \right ) \ .
\end{eqnarray}
Here $\mathcal{N}$ is a normalization constant. 

\begin{figure}[tb!]
\begin{center}
\scalebox{0.80}{\includegraphics{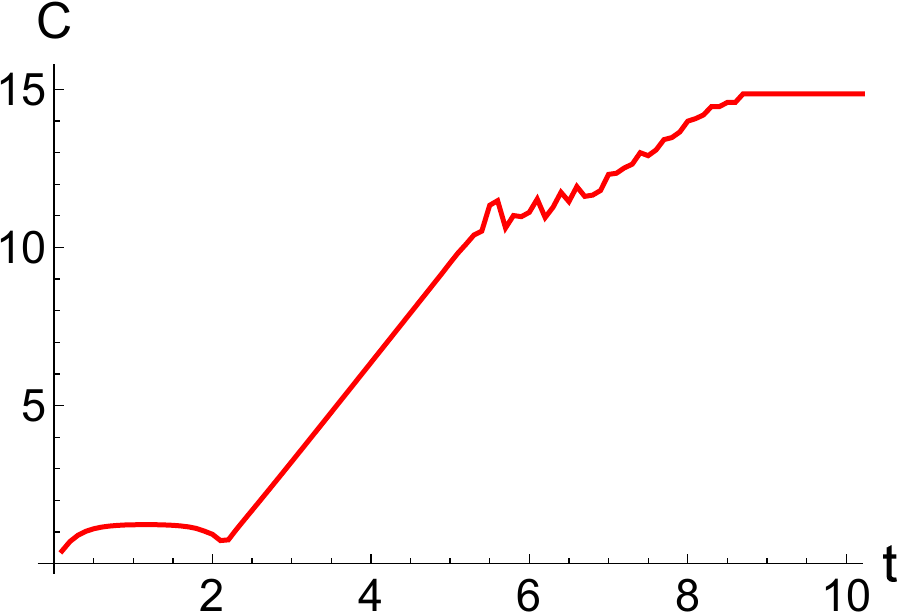} }
\scalebox{0.80}{\includegraphics{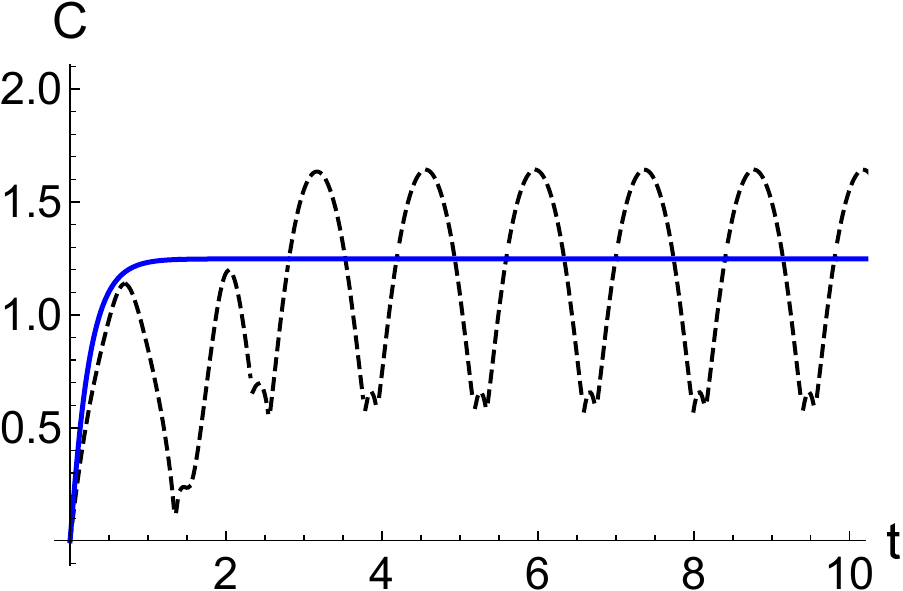} }
\vspace{-0.1in}
\caption{Complexity for the effective wave function when both oscillators are inverted (left panel) and when one of the oscillators are inverted. The parameters are -- {\bf Left:} $\epsilon_1= - 5.01, \epsilon_2= - 5, \lambda=0.1$ and {\bf Right:} black dashed curve has $\epsilon_1= 5.01, \epsilon_2=- 5, \lambda=0.1$  and the blue curve has $\epsilon_1= -5.01, \epsilon_2= 5, \lambda=0.1$.}
\label{sub1}
\end{center}
\end{figure}
The effective wave function can be written in the form
\begin{equation}
 \psi(x_1',x_1) = {\cal N}  \exp \left ( - \frac{1}{2} \left( \beta x_1^2 + \beta^* x_1'^2 \right)
      + \zeta  x_1 x_1' \right)  \ .
\end{equation}
To proceed, we need to diagonalize the argument of the exponential; we obtain the effective wave function
\begin{subequations}
\begin{equation}
 \psi(x',x) = {\cal N} 
  \exp \left (- \frac{1}{2} (\beta_1 + E) X_1^2 - \frac{1}{2} (\beta_1 - E) X_2^2 \right )
  \label{state2}
\end{equation}
where 
\begin{equation} 
 \left( \begin{array}{c}
    X_1 \\ X_2
    \end{array} \right) 
 = \left( \begin{array}{cc}
    u & -v \\ 
    v & u
    \end{array} \right) 
 \left( \begin{array}{c}
    x_1 \\ x'_1 
    \end{array} \right) 
    \label{waveeffective}
    \end{equation}
 and
 \begin{equation}
 E^2 = \zeta^2 - \beta_2^2 \ \ , \ \
 u = \sqrt{ \frac{1}{2} \left( 1 + i \ \frac{\beta_2}{E} \right) }
 \ , \ 
 v = - \sqrt{ \frac{1}{2} \left( 1 - i \ \frac{\beta_2}{E} \right) }  \ ,
\end{equation}
\end{subequations}
with $\beta_1 = {\rm Re}[\beta]$, $\beta_2 = {\rm Im}[\beta]$.
In what follows, we will use this effective wavefunction (\ref{waveeffective}) as the target state and compute the complexity. with respect to the ground state wavefunction.

Fig.~\ref{sub1} represents the complexity for this state (\ref{state2}). When both the system and bath are inverted ($\epsilon_1 <0, \epsilon_2<0$) the complexity for the effective wavefunction displays the expected chaotic like behaviour, namely it contains a small complexity scrambling period, followed by a linear growth and finally a saturation. On top of that, we see a downward concavity during the scrambling region. Note that unlike the full system complexity, when one of the oscillators is inverted, we do not see this behaviour as illustrated in the right panel of Fig.~\ref{sub1}. Understanding the physical meaning of this early time behaviour of complexity from the density matrix requires further investigation, and we would leave that for different work.
\begin{figure}[ht]
\begin{center}
\scalebox{1}{\includegraphics{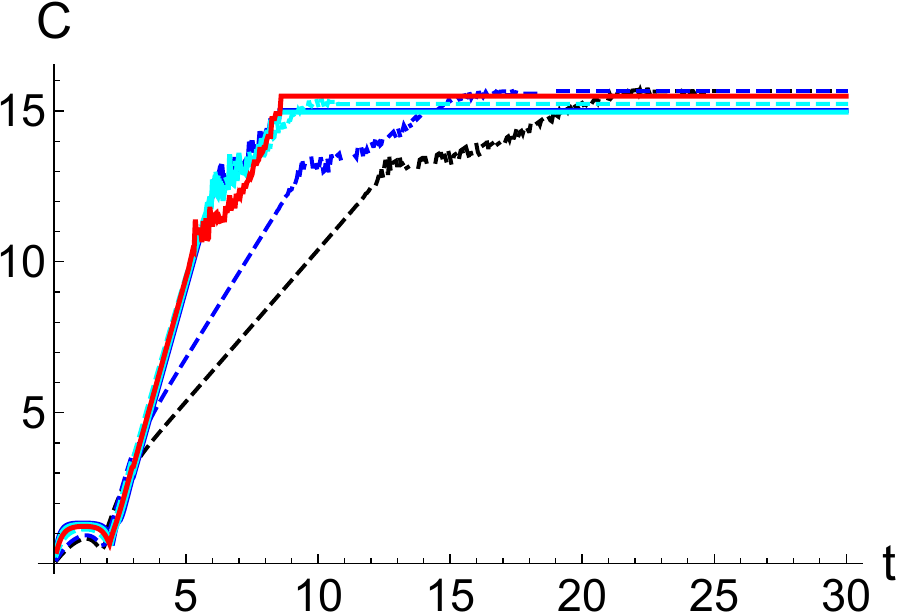} }
\scalebox{0.50}{\includegraphics{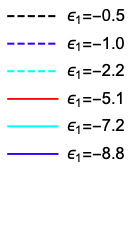} }
\vspace{-0.1in}
\caption{Complexity for $\epsilon_2=-5, \lambda=0.1$}
\label{sub2}
\end{center}
\end{figure}
Next, we want to investigate if this new diagnostic can sense the entire Lyapunov spectrum, as could the (full) system complexity. This is detailed in Fig.~\ref{sub2}. Below we list our findings: 

\begin{itemize}
    \item  When we fix the bath parameter $\epsilon_2 (<0)$ and gradually increase the inverted system parameter 
(starting with a smaller absolute value than the bath parameter), we see the same behaviour as the full-system complexity as shown in Fig.~\ref{all}. The linear portion is composed of two different slopes, just like the full system complexity. 

\item
Unlike the full system complexity, we only see \textit{one scrambling regime} when the bath parameter $\epsilon_2$ is kept fixed but the system parameter $\epsilon_1$ is being changed as shown in Fig.~\ref{sub2} \footnote{By the scrambling time we mean the time scale after which the complexity starts to grow linearly. It is evident from  Fig.~\ref{sub2}, that this scrambling time coincide with each other even if we change the system parameter $\epsilon_1$ for the fixed value of the bath parameter $\epsilon_2$. We are calling this as \textit{one scrambling regime.}}. 
\item
Furthermore, when the system and the bath parameters are close, the slopes (of the linear portion of the graph displayed in Fig.~\ref{sub2}) are aligned; it remains the same, even if the magnitude of the system parameter is larger than the bath parameter. It gives a bound on the growth rate of complexity (Lyapunov exponent), and this is controlled by the bath parameter. Curiously, this is consistent with the black hole case where the Lyapunov exponent is bounded by the bath temperature \cite{Maldacena_2016}. At this point it is only a qualitative comparison. We believe it deserves a more systematic study in future. {\sl This feature is not present in the full system case. }
\begin{figure}[ht]
\begin{center}
\scalebox{1}{\includegraphics{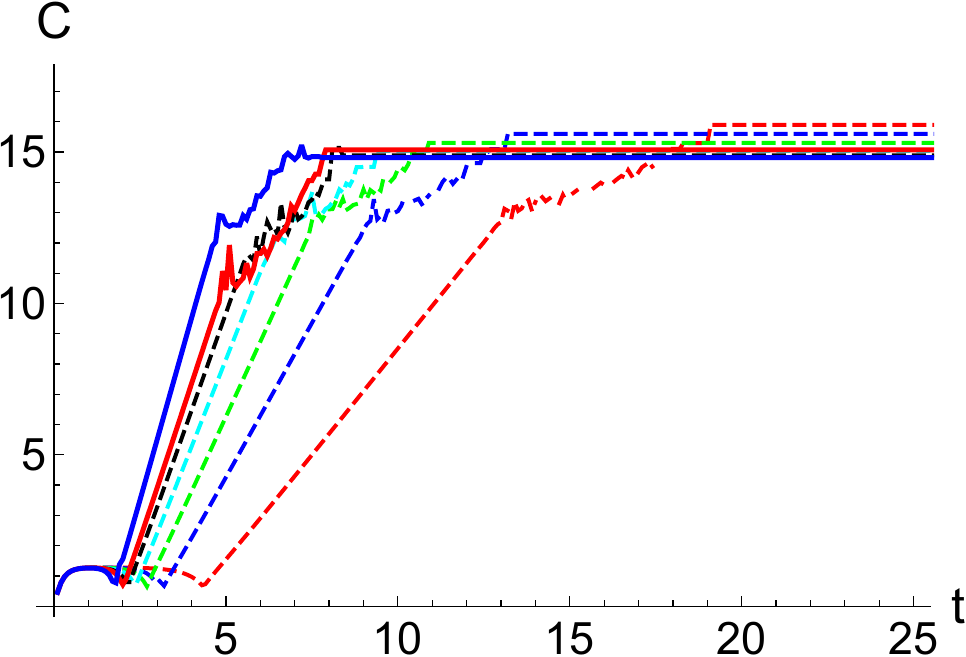} }
\scalebox{0.6}{\includegraphics{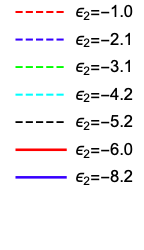} }
\vspace{-0.1in}
\caption{ Complexity with fixed $\epsilon_1=-6, \lambda=0.1$  }
\label{sub3}
\end{center}
\end{figure}
\item
We do not see the same scrambling regime if we fix the system parameter $\epsilon_1 (<0)$ and gradually increase the bath parameter (starting with a smaller absolute value than the system parameter) 
as in Fig.~\ref{sub3}. As we change the bath parameter, both the scrambling time and the Lyapunov exponent changes significantly.

\begin{figure}[ht]
\begin{center}
\scalebox{0.82}{\includegraphics{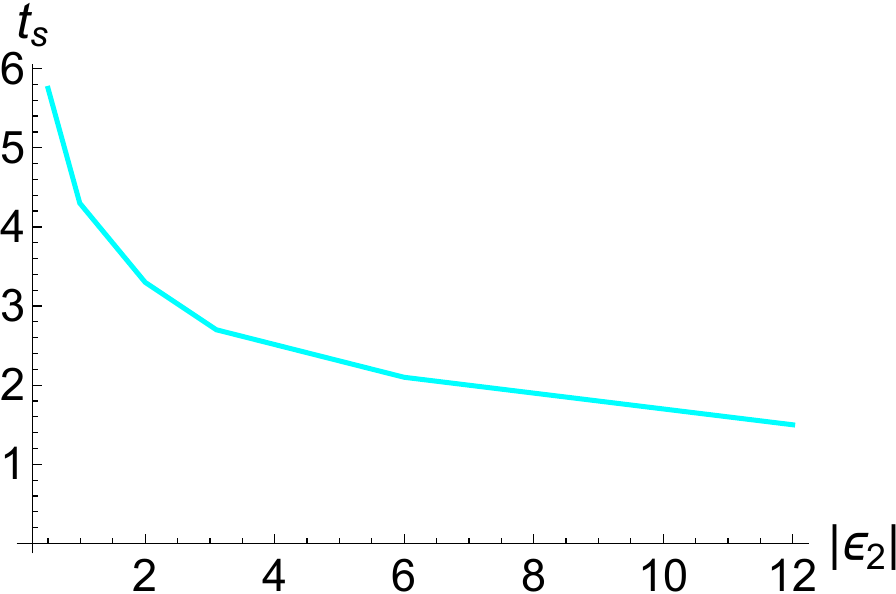}}
\scalebox{0.80}{\includegraphics{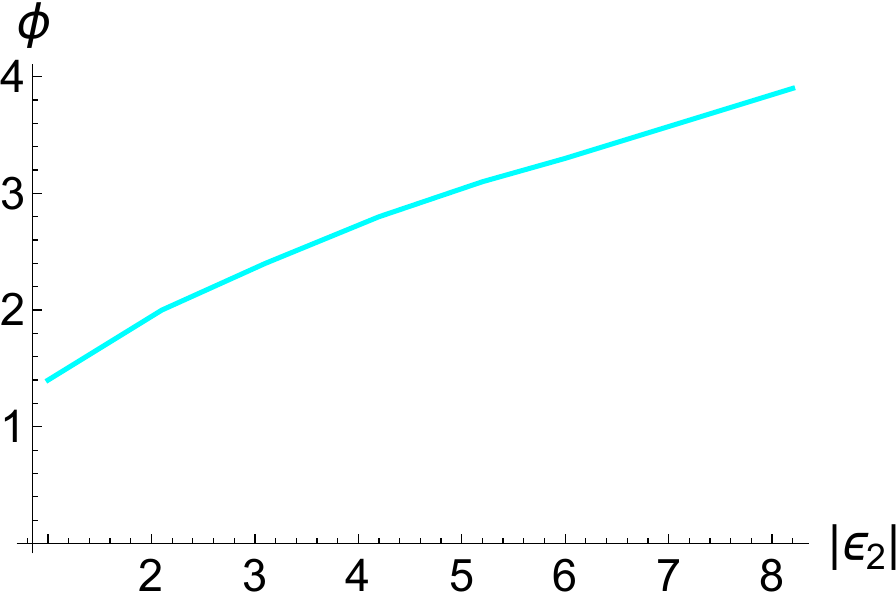} }
\end{center}
\vspace{-0.1in}
\caption{{\bf Left:} Scrambling time ($t_s$) vs $|\epsilon_2|$ , {\bf Right:} Slope ($\phi$) vs $|\epsilon_2|$}
\label{scram}
\end{figure}

\item We can explore this further by investigating how the scrambling time (time scale when complexity starts to grow linearly \cite{me3}) and the slope of the linear growth (Lyapunov exponent) change with changing the bath parameter keeping the system parameter and coupling fixed. The left panel of Fig.~\ref{scram} displays the change in scrambling time with bath parameter $(|\epsilon_2|)$ and the right panel displays the slope with bath parameter $(|\epsilon_2|)$. These figures indicate that as the bath gets more unstable/chaotic, the system scrambles faster, and the slope of the complexity grows bigger. Both these behaviours are consistent with the parameter dependence for single oscillator found in \cite{me3}.
\end{itemize}

\begin{figure}[ht]
\begin{center}
\scalebox{1}{\includegraphics{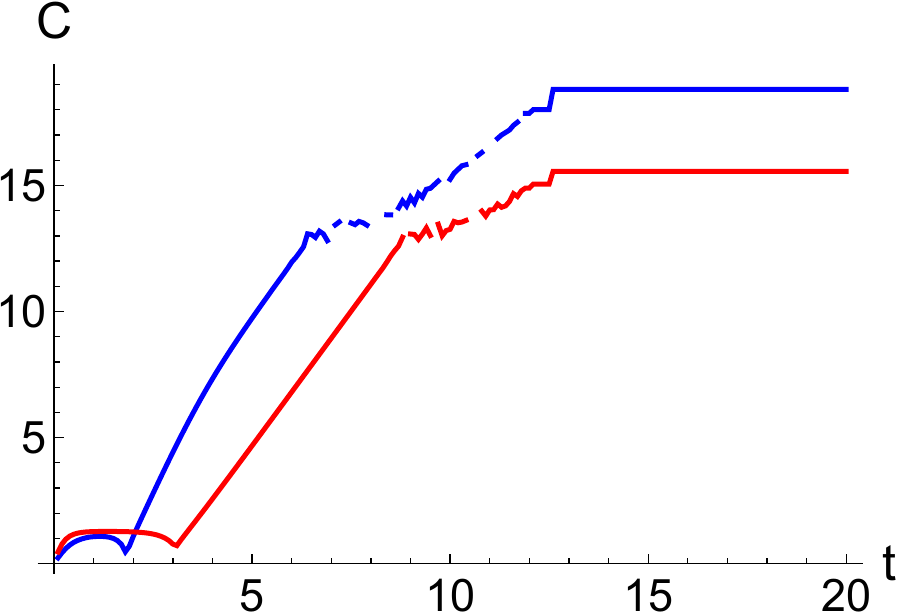} }
\vspace{-0.1in}
\caption{Blue: $\epsilon_1= -2.3, \epsilon_2= -6.1, \lambda=0.1$, \ Red: $\epsilon_1=-6.1 , \epsilon_2=-2.3, \lambda=0.1  $ }
\label{com}
\end{center}
\end{figure}

We will conclude this section by highlighting another difference from the full system case. Complexity for this effective wavefunction does not overlap with each other when the parameter $\epsilon_1$ and $\epsilon_2$ are switched as displayed in Fig.~\ref{com}. The underlying reason for this asymmetry is, of course, the tracing out of the bath oscillator (oscillator 2). This makes it more appropriate for the understanding of open systems. Since in all realistic scenario, we deal with open systems this complexity provides a more practical choice of investigating the underlying chaotic nature of the system. 


\subsection{Complexity of Purification}
To further illustrate the importance and the scope of our new diagnostic in this section we compute the complexity of purification \cite{Agn2019,Caceres:2019pgf,Camargo:2018eof,Camargo:2020yfv}, yet another method of computing complexity for mixed state. We will show that this particular approach is not as sensitive as complexity coming the operator-state mapping \cite{JAMIOLKOWSKI1972275,CHOI1975285,PhysRevA.87.022310} as discussed in the Sec.~(4.1).  

We start with the density (reduced) matrix defined in Eq. \ref{rho1}. First we purify ($|\psi_{11'}\rangle$) in such way that, 
\begin{equation} \label{cond}
\hat \rho_1= {\rm Tr}_{1'}|\psi_{11'} \rangle \langle \psi_{11'} |,
\end{equation}
$1'$ corresponds to the auxiliary Hilbert-space such that $|\psi_{11'}\rangle$ is a pure state.  Then the complexity of purification ($\mathcal{C}_p$) is defined as,
\begin{equation} \label{min}
\mathcal{C}_p= \min_{1'}\mathcal{C}(|\psi_{11'}\rangle),
\end{equation}
where, the minimization is over all possible purification and $\mathcal{C}(|\psi_{11'}\rangle)$ corresponds to the complexity of the state $|\psi_{11'}\rangle $ with respect to the reference state, which is the ground state of the Hamiltonian in Eq. \ref{Hsystem} with $\epsilon_1=\epsilon_2=\omega_0=1$ and $\lambda=0.$ Also, we choose to use minimal purification as advocated in \cite{Caceres:2019pgf,Bhattacharyya:2018sbw, Bhattacharyya:2019tsi} such that the size of the original (sub) system is same as the auxiliary system. In our case the subsystem is just one oscillator. \par
Next we parametrize our purified state in the following way, 
\begin{equation} \label{pur}
\psi(x_1,x_2)= \mathcal{N}\exp \left (-\frac{1}{2} \Big[\alpha x_1^2+\gamma x_2^2-2\, \tau x_1 x_2\Big]\right ),
\end{equation}
where $x_2$ belongs to the auxiliary Hilbert-space. $\alpha,\gamma $ and $\tau$ are in general complex and yet to be determined.  $\mathcal{N}$ is the normalization of the wavefunction. Then we have
\begin{equation}
\rho(x_1, x_2 \, | \, x_1', x_2')= \psi(x_1,  x_2) \psi^{*} (x_1', x_2')
\end{equation}
The corresponding reduced density matrix after tracing out the auxiliary Hilbert space is
\begin{align}
\begin{split}
&{\rm Tr}_{2}\rho(x_1, x_2 \, | \, x_1', x_2')\\&= \int_{-\infty}^{\infty} dx_2 \, \psi(x_1,  x_2) \psi^{*} (x_1', x_2)\\&
= \displaystyle{\mathcal{N}^2\exp\left (-\frac{1}{2}\left[\left(\alpha-\frac{\tau^2}{2\, {\rm Re}(\gamma)}\right)x_1^2+\left(\alpha^*-\frac{(\tau^*)^2}{2\, {\rm Re}(\gamma)}\right) x_1'^2\right]+\frac{|\tau|^2}{2\,{\rm Re}(\gamma)} x_1 x_1'\right)}
\end{split}
\end{align}
Using the condition mentioned in Eq. \ref{cond} and using Eq. \ref{rho1} we get the following,  
\begin{equation} \label{apr}
\alpha= \Omega_1(t), \tau= \kappa (t), {\rm Re }(\gamma)= {\rm Re }(\Omega_2(t)).
\end{equation}
\begin{figure}[t]
\begin{center}
\scalebox{0.80}{\includegraphics{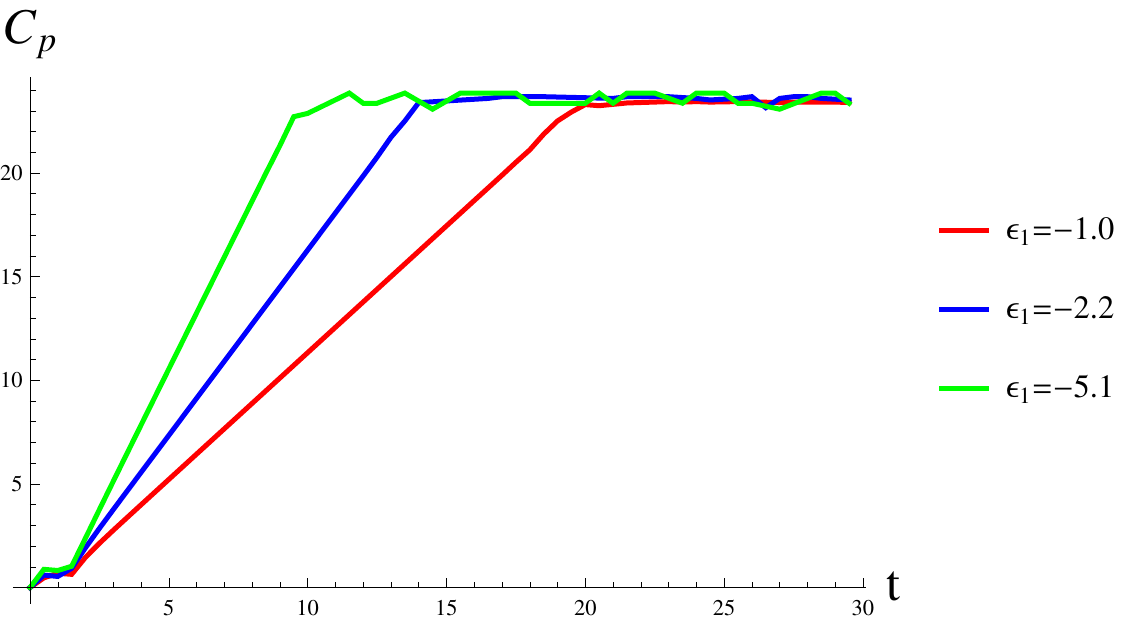} }
\scalebox{0.80}{\includegraphics{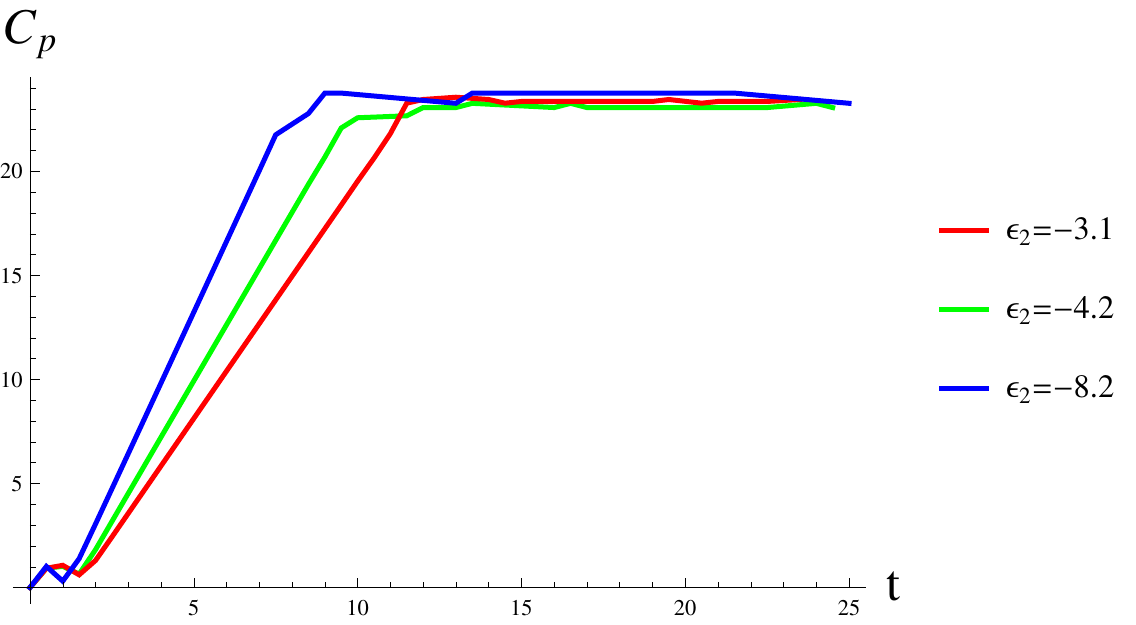} }
\vspace{-0.1in}
\caption{Evolution of Complexity of purification for $\lambda=0.1,$ {\bf Left:} $\epsilon_2=-5$ and $\epsilon_1$ is changed, {\bf Right:} $\epsilon_1=-6$ and $\epsilon_2$ is changed. }
\label{comf}
\end{center}
\end{figure}
We have determined all the parameters inside Eq. \ref{pur} in terms of the given parameters as in Eq. \ref{apr}, except for ${\rm Im }(\gamma)$. Hence the minimization in (\ref{min}) can be carried over  ${\rm Im }(\gamma)$ and the minimum value will correspond to the complexity of purification. We compute the complexity corresponding to Eq. \ref{pur} as follows:
\begin{equation}
\mathcal{C}(|\psi_{11'}\rangle)=\frac{1}{2} \sqrt{ \sum_{i=1}^{2} \left[
     \ln \left (\frac{| \omega_{i}|}{\omega_0} \right)^2 
  + \tan^{-1} \left (-\frac{ \text{Im} ( \omega_{i})}{\text{Re} ( \omega_{i})} \right)^2 
  \right] }  \ ,
\end{equation}
where, 
\begin{align}
\begin{split}
\omega_1=\frac{1}{2}(\alpha+\gamma+\sqrt{(\alpha-\gamma)^2+ 4\tau^2}), \ \omega_2=\frac{1}{2}(\alpha+\gamma-\sqrt{(\alpha-\gamma)^2+ 4\tau^2}),
\ \omega_0=1.
\end{split}
\end{align}
Finally, the complexity of purification is
\begin{equation}
\mathcal{C}_p=\min_{{\rm Im }(\gamma)}\Bigg(\frac{1}{2} \sqrt{ \sum_{i=1}^{2} \left[
     \ln \left (\frac{| \omega_{i}|}{\omega_0} \right)^2 
  + \tan^{-1} \left (-\frac{ \text{Im} ( \omega_{i})}{\text{Re} ( \omega_{i})} \right)^2 
  \right] } \Bigg).
\end{equation}

In Fig.~\ref{comf} (left panel) we display the evolution of this complexity for $\epsilon_2=-5,\ \lambda=0.1$ (as we did with the operator-state mapping \cite{JAMIOLKOWSKI1972275,CHOI1975285,PhysRevA.87.022310}).  We reach the following conclusions:
\begin{itemize}
    \item 
By comparing with Fig.~\ref{sub2}, we can see easily that $\mathcal{C}_{p}$ gives us a similar early time behaviour. This early time behaviour is then followed by a linear growth as expected for this kind of systems.
\item Unlike the complexity coming state-operator map, we only get a single slope, hence $\mathcal{C}_{p}$ doesn't give us the information about the Lyapunov spectrum. We can only extract one of the Lyapunov exponents. 
\item We have further explored the time-evolution of $\mathcal{C}_p$ with fixed $\epsilon_1$ and varying $\epsilon_2$ in Fig. \ref{comf} (right panel). But unlike Fig. \ref{sub2} and Fig. \ref{sub3} there is no asymmetry between the behaviour of  time evolution of $\mathcal{C}_p$ when we fix the system parameter and vary the bath parameter and vice versa.
\end{itemize}
In light of the above discussion, we would like to point out that the complexity of purification cannot detect the complete Lyapunov spectrum for this two oscillator model. \textit{Therefore, it seems it is not as sensitive as the complexity of mixed state obtained by using the operator-state mapping in capturing chaotic behaviour.} Finally, we would add that although we took the illustrative approach to explain our findings, the results outlined in this paper are robust as we have scanned the parameter space quite exhaustively.


\section{Discussion}

In this work, we took the first step toward using complexity to characterize chaos in a multiparticle system.  In the process, we took a step forward towards analyzing open quantum systems by applying the notion of complexity. Our model consisted of two oscillators, where one or both of the oscillators is inverted. Since the inverted oscillator is known to capture features similar to chaotic systems \cite {Blume-Kohout2003-ww}, this provides a natural toy model to study chaos. By exploring different types of quantum circuits, we showed that complexity is a useful diagnostic of chaos. 

We first considered the complexity of a doubly-evolved target state (forward followed by a backward evolutions with slightly different Hamiltonians)$| \psi_T \rangle = \exp(i H_>^B t) \exp( -i H_>^F t) | \psi_0 \rangle$, comparing the results with a singly-evolved target state $| \psi_T \rangle = \exp( -i H_> t) | \psi_0 \rangle$.
We showed that, for both the singly-evolved and doubly-evolved target states, the (full) system complexity exhibited different early-time behaviour, but are both bounded by the same saturation value.
When the parameters of the two oscillators are different, the linear growth region of the doubly-evolved state splits into two separate regions. Furthermore, the growth of complexity does not change when the parameters of the two oscillators are switched. 

Next, we proposed \textit{ a more natural diagnostic of chaos-based on the reduced density matrix} (where one of the particles is traced out) and the operator-state mapping \cite{JAMIOLKOWSKI1972275,CHOI1975285,PhysRevA.87.022310}. We discovered that the scrambling time and Lyapunov exponents are mainly dictated by the parameter of the particle that was traced out, i.e. by the `bath'. {\sl Qualitatively speaking, this is consistent with the black hole case where the Lyapunov exponent is bounded by the bath temperature} \cite{Maldacena_2016}.
We compared our results with those obtained using the complexity of purification; we showed that the complexity of purification gives less information, thus further highlighting the importance of the specific construction of our new proposal.

Besides its potential as a more natural testing device for chaotic behaviour, complexity from the density matrix has another advantage over the full system complexity. 
In this work, we considered an effective wave function $\psi(x,x') \sim \rho_1^{1/2}(x',x)$; this was motivated by the thermofield-double construction \cite{Hosur:2015ylk}.
However, one can consider effective wave functions built from more general powers of $\hat{\rho}_1$, $\psi(x,x') \sim \rho_1^{q}(x',x)$ --- 
this could provide further information by which to characterize the system; 
\color{black}
this proposal provides a more {\sl natural} form of complexity to compare with other information-theoretic measures such as Renyi entropy, entanglement entropy, etc., as they all are based on the same quantity, namely the reduced density matrix.  This might open up possibilities to explore complexity as an extension of entropy \cite{Brown:2017jil,Bernamonti:2019zyy} and other measures of correlations (for eg. OTOC) \cite{Syzranov:2018ikh,Tuziemski:2019rnx,Chakrabarty:2018dov} for open systems.

To give a proof-of-principle argument for complexity from the reduced density matrix as a new diagnostic for chaos, we have used the coupled inverted oscillators as a toy model. This is, however, a rather special example and, by no means, a realistic chaotic system. We want to explore this diagnostic for the realistic chaotic system in future work. Perhaps, a realistic model where one can apply this analysis is the chaotic spin chain model e.g. transverse field Ising model \cite{Hosur:2015ylk}. For that model one needs to find out geodesics on the space of $SU(N)$ unitaries, which we believe is tractable along the line of \cite{NL2, NL3}.

\section{Acknowledgments}
We would like to thank Aranya Bhattacharya for helpful conversations and discussions. A.B. is supported by  Start Up Research Grant (SRG/2020/001380) by Department of Science \& Technology Science and Engineering Research Board (India). S.H. would like to thank the University of Cape Town for funding this project. 

\end{document}